\documentclass[lettersize,journal]{IEEEtran}
\usepackage{amsmath,amsfonts}
\usepackage{algorithmic}
\usepackage{algorithm}
\usepackage{array}
\usepackage[caption=false,font=normalsize,labelfont=sf,textfont=sf]{subfig}
\usepackage{textcomp}
\usepackage{stfloats}
\usepackage{url}
\usepackage{verbatim}
\usepackage{graphicx}
\usepackage{cite}
\usepackage{graphicx}
\usepackage{cite}
\usepackage{xcolor}
\usepackage[normalem]{ulem}

\begin{document}

\title{Steady-State Statistical Modeling of Digitally Stabilized Laser Frequency with Markov-State Feedback}

\author{
Swarnav Banik$^{*,1}$, Elliot Greenwald$^{1}$, Xing Pan$^{1}$%
\thanks{$^{*}$swarnav@lightmatter.co}%
\thanks{$^{1}$Lightmatter Inc., 250 Bryant St, Mountain View, CA 94041}
\thanks{This work has been submitted to the IEEE for possible publication. Copyright may be transferred without notice, after which this version may no longer be accessible.}
}

\maketitle

\begin{abstract}
Laser frequency stabilization is conventionally analyzed using continuous-time control theory, which accurately models analog feedback but is insufficient for digital implementations where quantization, sampling, and stochastic noise shape the dynamics.
In modern digital laser systems, such as Photonic Integrated Circuit (PIC)-based lasers, finite discriminator and actuator resolution, sampling delays, and measurement noise introduce stochastic behavior that deterministic models do not capture.
We present a discrete-time Markov-state framework that models the evolution of the quantized actuator in a digital laser frequency lock, with state-transition probabilities determined by the frequency discriminator response, noise statistics, and implemented digital control logic.
The steady-state actuator and locked-laser frequency distributions are obtained directly from the unit-eigenvalue solution of the transition matrix, providing immediate access to key stability metrics without long time-domain simulations.
For white frequency noise, we show that the Markov formulation is exact under decorrelated sampling and update schemes, while correlated discriminator sampling introduces a predictable inflation of actuator variance without shifting the operating point.
In the presence of colored noise, long-range temporal correlations induce sampling-dependent deviations in both actuator mean and variance, defining the regime of validity of the memoryless Markov description.
This framework provides a compact and physically transparent tool for analyzing and optimizing digitally stabilized lasers in integrated photonic systems.
\end{abstract}

\begin{IEEEkeywords}
Digital Control Loops, Quantized Control Systems, Photonic Integrated Circuits (PICs)
\end{IEEEkeywords}

\section{Introduction}

Laser frequency stabilization is a critical requirement in modern optical and photonic systems, where precise wavelength control directly impacts system stability, capacity, and scalability.
In applications such as Dense Wavelength Division Multiplexing (DWDM) networks, coherent optical links, and emerging Artificial Intelligence-driven data-center interconnects, large numbers of closely spaced optical wavelengths must be stabilized simultaneously to maintain channel alignment, minimize crosstalk, and enable high spectral efficiency.
While traditional analog locking techniques, including Pound–Drever–Hall, optical beat-note locking, and modulation transfer spectroscopy, offer high control bandwidth and excellent noise performance \cite{drever1983, black2001}, many contemporary optical communication and sensing systems operate in regimes where ultra-high feedback bandwidth is unnecessary.
In these systems, digital control provides practical advantages, such as sharing hardware resources across channels and implementing diverse locking schemes in software, thereby scaling more efficiently than analog electronics \cite{zhang2024, idjadi2024}.

Unlike analog systems, digital feedback loops are discrete and stochastic. 
Sensing and actuation quantization, measurement noise, and sampling delay make the laser frequency evolve in discrete steps governed by probabilistic transitions. 
Sensor and actuator quantization can prevent the lock from reaching the exact target frequency, while measurement noise can induce random walk–like fluctuations or occasional loss of lock. 
Sampling delays introduce additional uncertainty by postponing corrective actions, which can lead to drift or oscillations near the lock point.
Consequently, stochastic state transitions and nonlinear behaviors arising from these effects, which we refer to as digital non-idealities, are not adequately captured by classical tools.
Furthermore, the specific choice of control loop design parameters, such as the employed dither scheme and update logic, influences the system's statistical behavior, producing stability properties that cannot be predicted by classical loop-gain or Bode analysis.
Related engineering fields have adopted probabilistic state-transition methods to model certain digital control systems \cite{sun2020, zhang2023, shen2018, hsu2015}.
However, to the best of our knowledge, these approaches have not been adapted for laser frequency stabilization, where semiconductor noise spectra, synchronous lock-in detection, and nonlinear optical discriminators produce state-dependent transition probabilities that differ from those in existing digital-control formulations.
Consequently, despite their growing prevalence, digital laser frequency stabilization methods lack a unified theoretical framework for predicting their steady-state behavior.
This gap limits our understanding of how digital non-idealities and control-loop design parameters influence performance, thereby constraining systematic optimization.

Here, we introduce a discrete-time statistical framework for analyzing digital laser frequency stabilization systems and establish its regime of validity.
We model the quantized actuator setpoint as a discrete-state variable whose evolution is governed by probabilistic transitions determined by the laser frequency noise, the discriminator response, and the implemented digital control logic.
When successive control updates depend only on the current actuator state and statistically independent noise realizations, the actuator dynamics satisfy the Markov property by construction.
Under these conditions, the feedback loop can be represented by a transition matrix whose unit-eigenvalue eigenvector yields the exact steady-state actuator probability distribution \cite{feller1968, norris97, ross2020}.
This distribution provides immediate access to key statistical properties of the lock, including residual frequency offset, steady-state variance, and convergence behavior, without requiring long time-domain simulations or extensive ensemble averaging.
It provides a transparent and computationally efficient means to quantify how digital non-idealities shape steady-state performance, thereby allowing efficient exploration of design trade-offs. 

Using this framework, we show that for white frequency noise the Markov prediction reproduces time-domain steady-state statistics, provided that the frequency discriminator sampling and control loop update schemes do not introduce inter-update memory.
This establishes the Markov description as a computationally efficient substitute for time-domain simulations in the memoryless regime.
We then use controlled deviations from this idealized case to identify how specific control loop design choices, such as finite-difference demodulation and update timing, introduce short-range temporal correlations that manifest as a systematic inflation of the actuator variance, even under white noise.
Finally, we extend the analysis to colored noise, where long-range temporal correlations violate the memoryless assumption and lead to quantitatively predictable departures between Markov and transient dynamics.
The remainder of this manuscript is organized as follows.
In Sec.~\ref{sec: arch} we introduce a representative digital laser frequency feedback loop and formulate its discrete-time dynamics as a Markov process.
Sec.~\ref{sec: sysModels} defines the free-running laser noise and introduces a universal frequency discriminator model that captures relevant hardware and control-loop design parameters.
Sec.~\ref{sec: whiteNoise} establishes exact agreement between the Markov and time-domain descriptions under white-noise-limited operation.
Sec.~\ref{sec: whiteNoiseDiscrepancy} examines controlled sources of variance inflation arising from correlated discriminator sampling.
Sec.~\ref{sec: coloredNoise} analyzes the breakdown of the Markov description in the presence of flicker noise.
Sec.~\ref{sec: conclusion} summarizes the implications and outlines future extensions.

\section{Discrete-Time Markov Representation of Digital Laser Frequency Locks}\label{sec: arch}

\begin{figure}[t!]
\centering
\includegraphics[width=\linewidth]{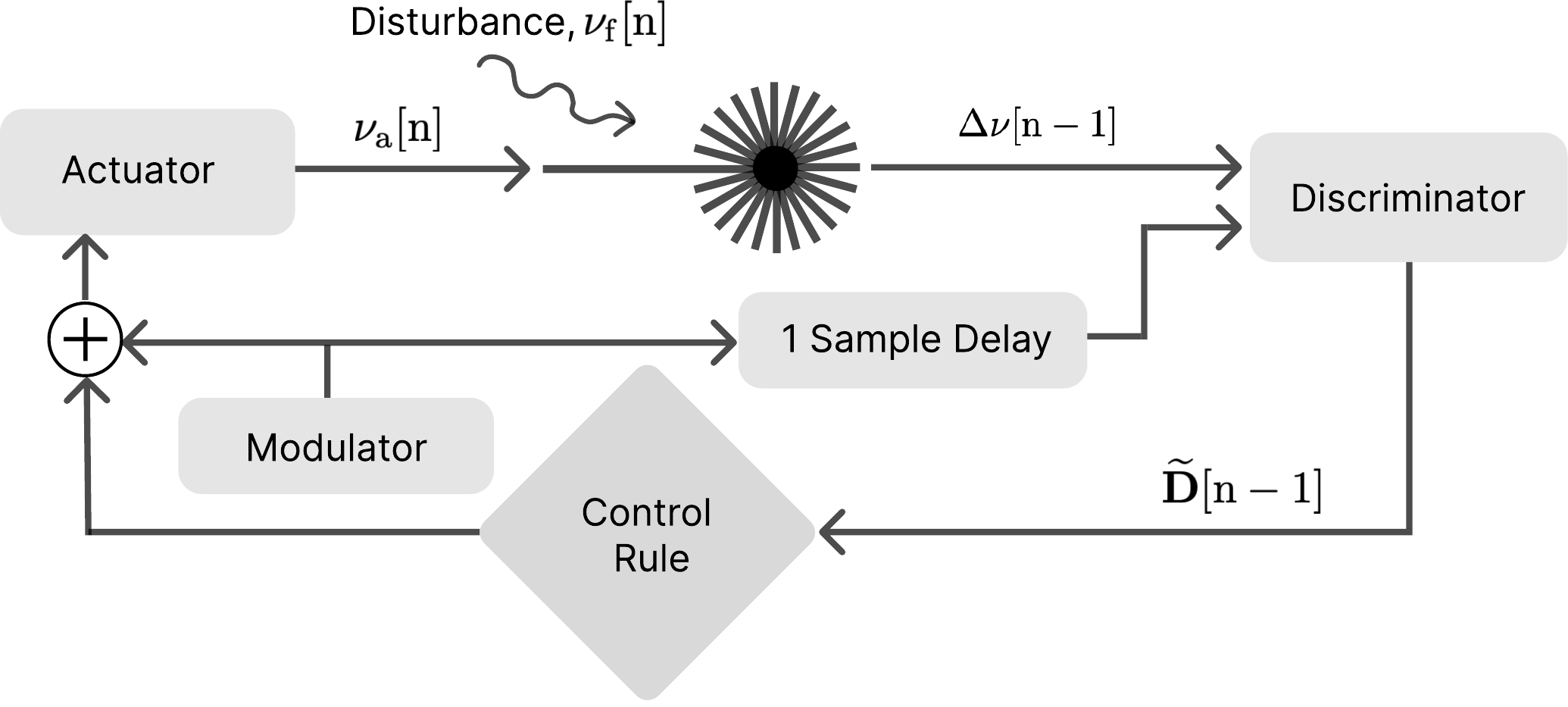}
\caption{Schematic of the digital laser frequency stabilization loop. 
The free-running laser frequency, $\nu_{\rm f}$, is compared against a frequency discriminator, whose output is digitized and synchronously demodulated using a discrete modulation sequence to generate a signed error signal, $\widetilde{\textbf{D}}$. 
A digital control rule evaluates the demodulated error at each sampling instant and updates a quantized actuator, realized by a finite-resolution DAC that applies a discrete frequency correction, $\nu_{\rm a}$, to the laser.
Measurement noise, quantization, and sampling discretization introduce stochasticity into the control decisions, resulting in probabilistic transitions between discrete actuator states.}
\label{fig: controlLoop}
\end{figure}

We model the digital laser frequency lock as a discrete-time stochastic process evolving over a finite set of actuator states, reflecting the quantized nature of the digitally controlled tuning element and its periodic update. 
Although the actuator itself is deterministic, stochastic transitions between discrete states arise from laser frequency noise and discriminator noise, which perturb the control decision at each update cycle.
A schematic of the digital feedback loop is shown in Fig.~\ref{fig: controlLoop}, illustrating the interaction between the laser, frequency discriminator, and the digital actuator.
At time index ${\rm n}$, the instantaneous frequency deviation from the discriminator reference is given by
\begin{equation}
\Delta \nu [{\rm n}] = \nu_{\rm f} [{\rm n}] + \nu_{\rm a}[{\rm n}],
\label{eq: freqStates}
\end{equation}
where $\nu_{\rm f}[{\rm n}]$ represents the stochastic free running laser optical frequency and $\nu_{\rm a}[{\rm n}]$ is the the actuator-applied frequency correction.
We model the actuator as a ${\rm N}_{\rm a}$-bit Digital-to-Analog Converter (DAC) whose output frequency correction is given by $\nu_{\rm a}[{\rm n}] = i \Delta \nu_{\rm a}$, where $i$ is the discrete actuator state index and $\Delta \nu_{\rm a}$ is the actuator frequency step size.
The state index $i$ takes values from the finite set $i \in S \equiv \{-2^{{\rm N}_{\rm a}-1}+1, \dots, 0, \dots, 2^{{\rm N}_{\rm a}-1}\}$.
To connect stochastic laser dynamics to discrete feedback updates, we require a frequency discriminator that maps the laser frequency deviation, $\Delta \nu$, to a measurable error signal, $\widetilde{\textbf{D}}$.
To avoid ambiguity in the feedback loop, this mapping should be monotonic over the range of laser frequency excursions expected during operation, ensuring that each discriminator output uniquely corresponds to a single laser frequency \cite{black2001, drever1983}.
Upon sampling $\widetilde{\textbf{D}}$, the controller applies a deterministic decision rule to increment, decrement, or hold the actuator state at the next update.
Although the control rule is deterministic, noise-induced fluctuations in the sampled discriminator output introduce uncertainty when the signal lies near a decision threshold.
Consequently, identical actuator states can yield different update outcomes at successive cycles, giving rise to probabilistic transitions between discrete actuator states.
These transition probabilities depend on the intrinsic laser frequency noise, digital non-idealities, and control-loop design parameters.
The actuator evolution satisfies the Markov property when the control decision at update cycle ${\rm n}$ depends only on the instantaneous frequency offset and the applied control rule, and not on prior actuator states, modulation dithers, or temporally correlated noise.
This condition is satisfied for white frequency noise when the discriminator sampling and update timing are chosen such that successive error signals are uncorrelated.
For a system satisfying the Markov memoryless property, we collect all state-to-state transition probabilities into a transition matrix $\mathbf{T}$, with entries ${\rm T}_{ij}$ normalized so that $\sum_{j \in S} {\rm T}_{ij} = 1$ for each $i\in S$. 
$\mathbf{T}$ advances the actuator state probability distribution, ${\rm P}_{\rm a}$, according to ${\rm P}_{\rm a}[{\rm n}+1] = \textbf{T}~{\rm P}_{\rm a}[{\rm n}]$.
As a result, the steady state actuator distribution, ${\rm P}_{\rm a, m}$, is given by the eigenvector problem 
\begin{equation}
{\rm P}_{\rm a, m} = \textbf{T}~{\rm P}_{\rm a, m},
\label{eq: eigenvalue}
\end{equation}
where ${\rm P}_{\rm a, m}$ represents the long-term probability of finding the actuator in each discrete state, directly quantifying the statistical properties of the control loop and the resulting lock stability.
Unlike time-domain simulations, which require explicit stepwise propagation and long averaging to capture rare transitions, the Markov approach yields these distributions immediately.
As a result, the computational cost of the Markov method scales with the number of discrete actuator states rather than the simulation duration or noise correlation time.

\section{Representative System Models}\label{sec: sysModels}

To evaluate and illustrate the Markov framework, we consider representative models of the free-running laser and the frequency discriminator.
We model the free running laser frequency as the sum of a frequency offset, $\nu_0$, and a residual noise, $\delta \nu$, according to $\nu_{\rm f} = \nu_0 + \delta \nu$.
We consider a semiconductor laser representative of diode and PIC-based devices, which typically exhibit a free-running Lorentzian linewidth, $\Delta \nu_{\rm lw}$, of 10 to 150 kHz and pronounced flicker noise at low frequencies \cite{antona2022, Guo2022, stern2020}.
To capture both the fast and slow stochastic fluctuations, we model the $\delta \nu$ as the sum of white and flicker components, with an ensemble-averaged single-sided Power Spectral Density (PSD)
\begin{equation}
{\rm S}_{\rm f}(f;~\Delta\nu_{\rm lw}, {\rm S}_0, \alpha) = \frac{\Delta\nu_{\rm lw}}{\pi} + \frac{{\rm S}_0}{f^\alpha},
\label{eq: freqNoise}
\end{equation}
where $f$ is the Fourier frequency.
While the flat term $\Delta\nu_{\rm lw}/\pi$ represents the fast white-noise contributions that dominate the short-term linewidth, the ${\rm S}_0/f^\alpha$ flicker term captures slower, low-frequency drifts.
For our simulations, we set $\alpha = 1$ and explore experimentally observed ranges of white and flicker noise by varying $\Delta \nu_{\rm lw}$ and ${\rm S}_0$ \cite{rønnekleiv2001, martin-sanchez2024, jin2021, antona2022, Guo2022, stern2020}.
Appendix~\ref{apdx: laserNoise} summarizes the steps for obtaining numerical realizations of $\delta \nu$.

To ensure unambiguous control, the discriminator error signal, $\widetilde{\textbf{D}}$, should be monotonic over the range of frequency excursions encountered by the laser.
This can be achieved by a broad class of photonic frequency sensors, such as ring resonators, arrayed waveguide gratings (AWGs), Mach–Zehnder interferometers, and integrated Fabry–Pérot cavities, which typically exhibit a response that is symmetric about the resonance frequency.
While resonators such as microrings or Fabry–Pérot cavities typically have a Lorentzian-shaped frequency response, AWGs and distributed Bragg reflectors often exhibit Gaussian broadening due to fabrication disorder and thermal or carrier fluctuations, which smear the Lorentzian peak into a Gaussian profile.
The Voigt profile, which smoothly interpolates between Lorentzian and Gaussian limits, therefore provides a convenient and widely applicable representation of these diverse photonic discriminators \cite{yariv2007, little1997, okamoto2014}.
Accordingly, we model the raw discriminator output as a discretized Voigt function,
\begin{equation}
\textbf{D}(\Delta \nu) = {\rm V}_{\rm Q}(\Delta \nu;~\mathnormal{w},\gamma,\Delta \textbf{D}, \sigma_{\rm d}),
\label{eq: discRawSignal}
\end{equation}
where $\{\mathnormal{w},\gamma\}$ are the Gaussian and Lorentzian width parameters, $\Delta \textbf{D}$ is the Analog to Digital Converter (ADC) quantization step, and $\sigma_{\rm d}$ is the Root Mean Squared (RMS) sensor noise, which is modeled as an additive white, uncorrelated noise.
The function ${\rm V}_{\rm Q}$ accounts for the analog lineshape, additive white sensor noise, and digital discretization.

To obtain a monotonic, signed error signal for feedback, we implement a dither-based synchronous modulation-demodulation scheme.
Unlike conventional linearized analyses that approximate the frequency discriminator by a constant small-signal gain, we explicitly model the discrete modulation–demodulation process consistent with practical digital implementations\footnote{Here, modulation is not used to encode information, but to generate a monotonic, signed estimate of the discriminator slope at each control update.}.
At each sampling instant ${\rm n}$, the demodulated error signal $\widetilde{\mathbf{D}}$ is computed by correlating the discriminator output with the applied modulation sequence.
For discrete dither patterns, this correlation reduces to
\begin{equation}
\widetilde{\mathbf{D}}[{\rm n}] = \Delta {\rm M}~\left(\mathbf{D}[{\rm n}] - \mathbf{D}[{\rm n}-1]\right),
\label{eq: discErrorSignal}
\end{equation}
where $\Delta {\rm M}$ is the applied modulation change between samples.
This method, widely used in digital and integrated photonic systems, yields an error proportional to the local slope ${\rm d}\mathbf{D}/{\rm d}\nu$, making the lock robust to slow drifts and offsets \cite{rubiola2008}.
While the laser could be locked to any fixed error value, in this manuscript, we lock our lasers to the resonance frequency, such that ideal stabilization results in $\widetilde{\mathbf{D}}=0$.
Appendix~\ref{apdx: discriminator} provides example frequency response for some Voigt discriminators.
Later, in Sec.~\ref{sec: whiteNoiseDiscrepancy}, we show that the differencing operation, Eq.~\ref{eq: discErrorSignal}, introduces short-range temporal correlations, leading to a predictable variance inflation that rescales the actuator and locked-laser frequency statistics.
Because linearized discriminator models remove the modulation–demodulation pathway, they fundamentally cannot capture this variance inflation.

\section{White-Noise Validation of the Markov Framework}\label{sec: whiteNoise}

We first validate the Markov framework in the white-noise regime under a decorrelated discriminator sampling scheme that enforces statistical independence between successive control updates.
This configuration satisfies the memoryless assumptions and provides a baseline against which we assess the effects of correlated discriminator sampling and colored noise in later sections.
In the following subsections, we demonstrate that, under these conditions, the Markov model accurately reproduces both the steady-state and transient statistics of a full time-domain simulation.

\subsection{Decorrelated Discriminator Sampling \& Control Update}

To eliminate inter-update memory, we adopt a return-to-zero (RZ) modulation–demodulation scheme in which each nonzero dither pulse is followed by a zero-valued sample, and control updates are applied only on the zero cycles.
Although the demodulated error in Eq.~\ref{eq: discErrorSignal} is formed as a finite difference of two discriminator samples, restricting updates to the zero cycles prevents reuse of discriminator noise across consecutive control decisions, such that successive actuator updates depend on statistically independent noise realizations.
Additionally, we randomize the polarity of the nonzero dither pulses between $+A$ and $-A$, with $A = 40$ kHz, to avoid introducing a deterministic temporal structure in the sampled error signal.
To guide state transitions, we use the sign-based control rule:
\begin{equation}
i[{\rm n}+1]= 
\begin{cases}
    \text{min}\left(i[{\rm n}]+1,~i_{\rm max}\right),& \text{if  } \widetilde{\textbf{D}}[{\rm n}]> 0\\
    i[{\rm n}],& \text{if  } \widetilde{\textbf{D}}[{\rm n}]= 0\\
    \text{max}\left(i[{\rm n}]-1,~i_{\rm min}\right),& \text{if  } \widetilde{\textbf{D}}[{\rm n}]< 0
\end{cases}
\label{eq: controlLogic}
\end{equation}
so that the system moves in the direction of zero error.
Here, $i_{\rm min}$ and $i_{\rm max}$ correspond to the extreme values of the discrete set $S$.
Although the control loop includes quantization, synchronous demodulation, and nonlinear decision logic, we find that enforcing decorrelated discriminator sampling is sufficient for a memoryless Markov description to remain quantitatively accurate, reproducing time-domain steady-state statistics and convergence behavior.

\subsection{Estimating the Markov Prediction}

\begin{figure}[t!]
\centering
\includegraphics[width=\linewidth]{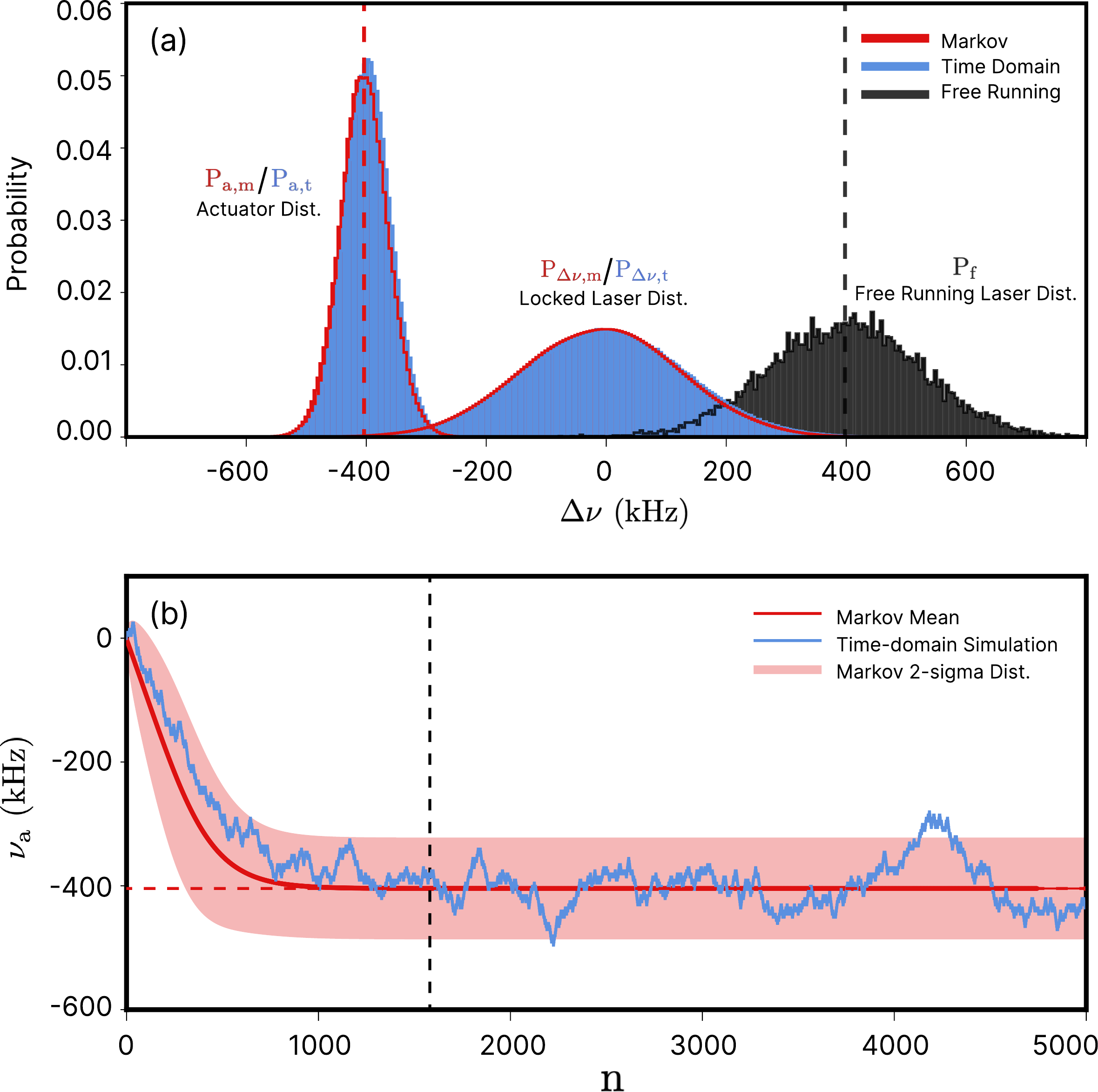}
\caption{Description of the Markov method and comparison with time-domain simulations. (a) Steady-state probability distributions of the free-running laser detuning (gray), actuator state, and locked-laser detuning obtained via the Markov method (red) and time-domain simulations (blue). Since the blue histograms obscure the red histograms, we draw the outline of the red histograms on top of the blue ones. The dashed vertical lines indicate the means of the free running laser frequency and Markov actuator distributions. (b) Time evolution of the actuator state as predicted by the Markov propagation and that observed in a single time-domain realization, illustrating convergence to steady state. The horizontal dashed red line denotes the mean of the ${\rm P}_{\rm a, m}$ distribution and the vertical dashed gray line marks ${\rm n}_{\rm conv}$.}
\label{fig: markov}
\end{figure}

We compute the transition matrix $\mathbf{T}$ and the resulting actuator steady-state distribution $\mathrm{P}_{\rm a,m}$, when sensed with a discrete Voigt discriminator and corrected using an actuator with step size $\Delta\nu_{\rm a} = 5$ kHz.
We generate white optical frequency noise by setting ${\rm S}_{0} =0$ and $\Delta \nu_{\rm lw} = $ 100 kHz in Eq.~\ref{eq: freqNoise}, and using the resulting frequency fluctuations, $\delta \nu$.
We consider a Voigt discriminator with widths $(\gamma,~\mathnormal{w}) = (1,~2.5)$ MHz, a quantization step $\Delta \textbf{D} = \textbf{D}_{\rm span}/2^{12}$, and additive sensor noise $\sigma_{\rm d} = 10^{-5}~\textbf{D}_{\rm span}$ which is an order of magnitude smaller than the quantization step. 
Here, $\textbf{D}_{\rm span}$ denotes the full dynamic range of the raw discriminator signal.
To highlight frequency tracking within the monotonic region of the discriminator error signal, we set the laser frequency offset, $\nu_0 = 400$ kHz.
While actuator states near resonance have comparable probabilities of transitioning in either direction, the states far from resonance have a negligible probability of transitioning away from the lock point.
Therefore, to limit the size of the transition matrix, $\mathbf{T}$, we truncate the range of actuator states by choosing $i_{\rm min}$ and $i_{\rm max}$ such that $|i~\Delta\nu_{\rm a}|<1$ MHz.
For our chosen parameters, the probabilities of the actuator reaching states beyond these bounds are less than $10^{-12}$.
As a result, this truncation reduces computational cost without affecting the conclusions\footnote{Our steady state distributions converge to stable solutions for $i_{\rm max/min}~\Delta\nu_{\rm a}<\pm 1$ MHz. 
Beyond this range, further expanding the allowable actuator indices has no effect on the steady-state distribution ${\rm P}_{\rm a, m}$, indicating that the chosen limits are already sufficiently wide to capture all relevant dynamics.}.
We sample $10^4$ time-samples of $\delta \nu$ and plot the sampled frequency noise probability distribution ${\rm P}_{\rm f}$,  as the gray bars in Fig.~\ref{fig: markov}(a)\footnote{To ensure consistency across the probability distributions, we bin all histograms into bin sizes equal to the actuator step size, $\Delta \nu_{\rm a}$.}.
For each actuator state $i$, we determine the transition probabilities ${\rm T}_{ij}$ by calculating the discriminator response for all the $10^4$ stochastic realizations of $\delta \nu$ and applying the control rule in Eq.~\ref{eq: controlLogic}.
In the limit of frequency excursions smaller than the linear regime of the discriminator response, all actuator states would experience identical error-signal statistics. 
The full transition matrix could therefore be generated from a single error-distribution evaluation, reducing the computational cost by orders of magnitude. 
Because our study spans a wider monotonic region, we retain the more general state-dependent formulation of $\textbf{T}$ to ensure accuracy across the entire operating range.
After estimating $\textbf{T}$, we solve Eq.~\ref{eq: eigenvalue} to obtain the steady state actuator distribution, ${\rm P}_{\rm a, m}$, plotted as red bars in Fig.~\ref{fig: markov}(a).

The resulting distributions reveal how the feedback loop suppresses slow drifts while leaving high-frequency fluctuations largely unaltered.
${\rm P}_{\rm a, m}$ and ${\rm P}_{\rm f}$ exhibit means of roughly equal magnitude and opposite sign, as shown by the gray and red dashed vertical lines.
This demonstrates that the feedback loop counteracts slow drifts in the laser frequency by producing an anti-correlation between the offset $\nu_0$ and the actuator state mean, $\langle \nu_{\rm a} \rangle$.
It allows us to approximate Eq.~\ref{eq: freqStates} to $\Delta \nu \approx \delta \nu + \widetilde{\nu}_{\rm a}$, where $\widetilde{\nu}_{\rm a} = \nu_{\rm a} - \langle \nu_{\rm a}\rangle$.
For low-bandwidth locks with white optical frequency noise, the residual actuator fluctuations, $\widetilde{\nu}_{\rm a}$, are effectively independent of the free-running noise, $\delta \nu$.
Under this approximation, the probability distribution of the instantaneous laser frequency, ${\rm P}_{\Delta \nu}$, can be expressed as the convolution of the distributions of $\delta \nu$ and $\widetilde{\nu}_{\rm a}$.
Since these distributions are shifted versions of ${\rm P}_{\rm f}$ and ${\rm P}_{\rm a}$ by roughly equal amounts in opposite directions, we can express ${\rm P}_{\Delta \nu}(\Delta \nu) \approx \left( {\rm P}_{\rm f} * {\rm P}_{\rm a}\right) (\Delta \nu)$.
Consequently, we determine the steady state frequency distribution of the locked laser, ${\rm P}_{\Delta \nu, \rm m}$, according to
\begin{equation}
{\rm P}_{\Delta \nu, \rm m} = \left( {\rm P}_{\rm f} * {\rm P}_{\rm a, m}\right) (\Delta \nu),
\label{eq: convl}
\end{equation}
and plot it as red bars in Fig.~\ref{fig: markov}(a), which, as expected, is roughly centered around zero. 
Eq.~\ref{eq: convl} and the resulting ${\rm P}_{\Delta \nu, \rm m}$ distribution provide an intuitive picture: the actuator suppresses slow drifts, while high-frequency noise is governed by the intrinsic laser noise, digital non-idealities, and the control loop design parameters.
While the example above uses a sign-based control rule (Eq.~\ref{eq: controlLogic}), single-difference demodulation (Eq.~\ref{eq: discErrorSignal}), and a randomized RZ dither, the Markov framework is not limited to these choices.
Alternative control strategies, such as proportional or multi-bit rules, as well as alternate dither schemes, can be incorporated by appropriately adjusting the transition probabilities $\mathbf{T}$.
This flexibility ensures that the framework can capture the steady-state statistics of a wide range of digital control loops, providing a general tool for analyzing the impact of different loop designs on actuator behavior and locked-laser frequency distributions.

\subsection{Comparison of Markov and Time-Domain Simulations}

To validate the discrete-time Markov framework, we compare its predicted actuator dynamics with those obtained from full time-domain simulations of the digital feedback loop. 
While the steady-state actuator distribution follows directly from the unit-eigenvalue solution of the transition matrix (Eq.~\ref{eq: eigenvalue}), step-wise propagation of the actuator probability distribution provides a direct means of comparing convergence behavior and transient evolution against time-domain trajectories. 
Starting from an initial actuator state localized at $i = 0$, ${\rm P}_{\rm a,0}$, we propagate the distribution according to ${\rm P}_{\rm a}[{\rm n}+1] = \textbf{T}~{\rm P}_{\rm a}[{\rm n}]$, such that ${\rm P}_{\rm a}[{\rm n}] = \textbf{T}^{\rm n}~{\rm P}_{\rm a,0}$.
The resulting evolution of the mean actuator value and its spread is shown in red in Fig. \ref{fig: markov}(b), where the shaded region denotes the two-sigma interval. 
We define the convergence time , ${\rm n}_{\rm conv}$, as the smallest iteration for which the total variation distance (L1 norm) to the steady-state distribution satisfies $||{\rm P}_{\rm a}[{\rm n}]- {\rm P}_{\rm a,m}||_1 < 10^{-3}$ \cite{levin2006}.
For the time-domain simulations, we initialize the actuator in state ${\rm P}_{\rm a, 0}$, impart stochastic frequency noise according to Eq.~\ref{eq: freqNoise}, and detect the resulting laser frequency offset using the Voigt discriminator.
We then update the actuator using the digital control rule in Eq.~\ref{eq: controlLogic} for $10^6$ cycles.
The blue trace in Fig.~\ref{fig: markov}(b) shows the evolution of $\nu_{\rm a}$ for the first 2500 cycles, demonstrating close agreement with the Markov-predicted convergence time and steady-state mean.
We extract the actuator values for ${\rm n}>{\rm n}_{\rm conv}$, and estimate the corresponding steady-state actuator distribution, ${\rm P}_{\rm a,t}$, and the instantaneous locked laser frequency distribution, ${\rm P}_{\Delta\nu,\rm t}$\footnote{The subscripts ${\rm m}$ and ${\rm t}$ denote Markov prediction and time-domain results}, and plot these distributions as blue bars in Fig.~\ref{fig: markov}(a).
The mean actuator values predicted by the Markov model and the time-domain simulations, agree to within 1 \% of the free-running frequency offset $\nu_0$, while their steady-state standard deviations differ by 3 \%, demonstrating that both approaches converge to the same operating point and reproduce the same steady-state fluctuations.

\begin{figure}[t!]
\centering
\includegraphics[width=\linewidth]{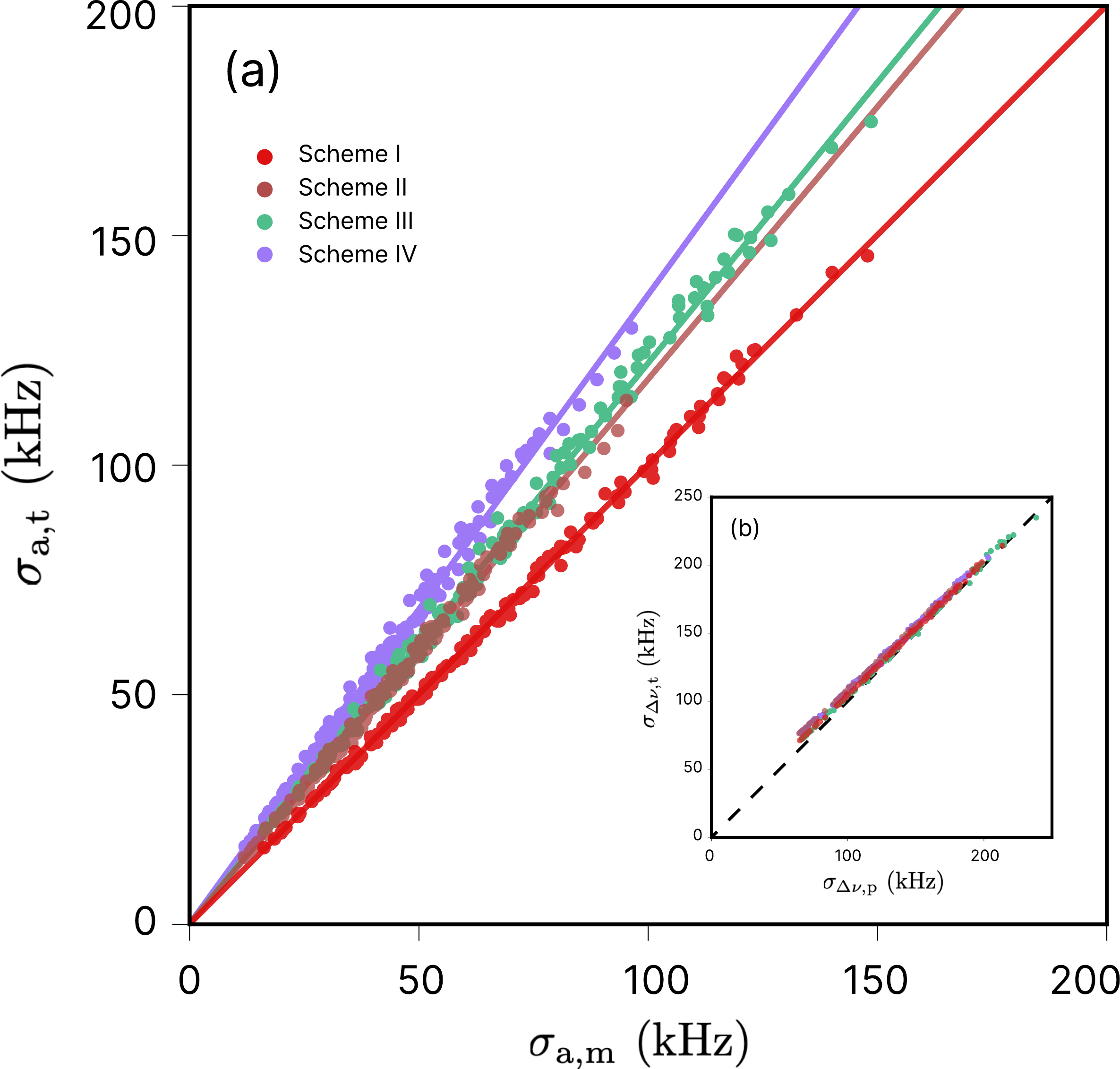}
\caption{Comparison of Markov and time-domain simulation variances for the schemes in Table~\ref{tab: correlationSchemes}. (a) $\sigma_{\rm a,t}$ vs. $\sigma_{\rm a,m}$. The solid lines are Eq.~\ref{eq: broadeningActuator} fitted to the data. (b) $\sigma_{\Delta \nu,t}$ vs. $\sigma_{\Delta \nu,p}$. Dashed black line is a guide line representing unity slope.}
\label{fig: discrepancy}
\end{figure}

To systematically assess the agreement between the Markov predictions and time-domain simulations, we perform a broad sweep over the parameters governing the actuator dynamics.
The free-running laser linewidth, $\Delta \nu_{\rm lw}$, and the discriminator noise, $\sigma_{\rm d}$, set the stochastic forcing and therefore dominate the width of the steady-state actuator distributions.
We vary $\Delta \nu_{\rm lw}$ and $\sigma_{\rm d}$ to access regimes in which either intrinsic laser noise or discriminator noise sets the dominant stochastic forcing.
To ensure that the conclusions do not depend on the discriminator shape or the control resolution, we also vary the discriminator Gaussian width, $\mathnormal{w}$, and the actuator step size, $\Delta \nu_{\rm a}$.
Specifically, we consider all permutations of the parameters listed in Table~\ref{tab: parameterSweep}, while restricting the laser frequency noise to be purely white.
\begin{table}[]
\centering
\caption{Parameter sweep for comparing Markov predictions and time-domain simulations.}
\label{tab: parameterSweep}
\begin{tabular}{ll}
\hline
\hline
\textbf{Parameter} & \textbf{Values} \\
\hline

$\Delta \nu_{\rm lw}$ 
& $\{25,\,50,\,75,\,100,\,125,\,150\}$ kHz \\[4pt]

$\sigma_{\rm d}$ 
& $\{10^{-3},\,10^{-4},\,10^{-5}\}\times\,\textbf{D}_{\rm span}$ \\[4pt]

$\mathnormal{w}$ 
& $\{2.5,\,5,\,10\}$ MHz \\[4pt]

$\Delta \nu_{\rm a}$ 
& $\{5,\,10,\,20\}$ kHz \\
\hline
\hline
\end{tabular}
\end{table}
Across all parameter combinations, the difference between the Markov-predicted and time-domain actuator means, $|\Delta \mu_{\rm a}|$, is, on average, only $0.06\%$ of the free-running laser offset $\nu_0$, confirming that both approaches converge to the same locked operating point.
This parameter sweep produces actuator and locked-frequency standard deviations spanning more than an order of magnitude, as shown in Fig.~\ref{fig: discrepancy}.
For each case, we compute the steady-state actuator standard deviations predicted by the Markov model, $\sigma_{\rm a,m}$, and obtained from time-domain simulations, $\sigma_{\rm a,t}$, along with the corresponding locked-frequency values, $\sigma_{\Delta\nu,\rm m}$ and $\sigma_{\Delta\nu,\rm t}$.
We plot $\sigma_{\rm a,t}$ as a function of $\sigma_{\rm a,m}$ as the red points in Fig.~\ref{fig: discrepancy}(a), where the data cluster tightly around the identity line $\sigma_{\rm a,t}=\sigma_{\rm a,m}$.
Fitting the data to 
\begin{equation}
\sigma_{\rm a, t} = \kappa~\sigma_{\rm a, m},
\label{eq: broadeningActuator}
\end{equation}
gives $\kappa = 1.000$ with a 1-sigma fit uncertainty of 0.001, demonstrating near-perfect quantitative agreement.
We compare the time-domain locked-frequency standard deviation, $\sigma_{\Delta \nu,\rm t}$, with the corresponding Markov-based prediction,
\begin{equation}
\sigma_{\Delta \nu, \rm p} = (\kappa^2~\sigma_{\rm a, m}^2+\sigma_{\rm f}^2)^{1/2},\\
\label{eq: adjustedMarkov}
\end{equation}
where $\sigma_{\rm f}^2$ is the variance of the free running frequency distribution, ${\rm P}_{\rm f}$, in Fig.~\ref{fig: discrepancy}(b).
The data clusters around the dashed $\sigma_{\rm \Delta \nu ,t}=\sigma_{\rm \Delta \nu,p}$ line.

\section{Correlated Discriminator Sampling \& Variance Inflation}\label{sec: whiteNoiseDiscrepancy}

The exact agreement demonstrated in Sec.~\ref{sec: whiteNoise} relies on a deliberately constructed sampling and update scheme that eliminates inter-update memory. 
While this provides a clean baseline, many practical digital laser locks update the actuator using discriminator error signals derived from correlated samples. 
In such implementations, successive control decisions reuse overlapping discriminator measurements, violating the strictly memoryless assumption of the Markov construction.
Even for purely white frequency noise, the finite-difference demodulation in Eq.~\ref{eq: discErrorSignal} introduces short-range temporal correlations in the sampled error signal. 
In this section, we show that these correlations lead to a systematic inflation of the steady-state actuator and locked-laser frequency variances.
This variance inflation arises entirely from discrete-time implementation choices, such as modulation format, demodulation method, and update timing, and is therefore not captured by linearized or small-signal models that abstract away the modulation–demodulation pathway.

To systematically explore the effects of correlated discriminator sampling, we consider three representative modifications to the baseline decorrelated scheme, summarized in Table~\ref{tab: correlationSchemes}.
In all three  (Schemes~II, III, and IV), the actuator is updated on every modulation cycle, so that successive control decisions reuse discriminator samples and introduce inter-update memory.
Scheme~III retains the randomized RZ modulation of Scheme~I but removes update gating, such that correlations arise solely from overlapping samples in the finite-difference demodulation.
Scheme~II removes the RZ padding while retaining randomized dither.
The random dither polarity breaks phase coherence between successive cycles, resulting in weaker temporal correlations and reduced variance inflation.
Scheme~IV replaces the randomized dither with a deterministic alternating-sign modulation sequence, in which the dither alternates between $+A$ and $-A$ on successive cycles, producing the strongest correlations due to the combined effects of overlapping samples and periodicity.
For each scheme, we repeat the same parameter sweep used for Scheme~I, compute the difference between the Markov-predicted and time-domain actuator means, and plot the resulting variances in Fig.~\ref{fig: discrepancy}(a).
As in the decorrelated case, the Markov and time-domain actuator means agree to within 1\% of the free-running laser offset $\nu_0$ for all three correlated schemes, confirming that correlated sampling does not shift the locked operating point.
However, the actuator variances lie along straight lines through the origin with $\sigma_{\rm a,t} > \sigma_{\rm a,m}$, indicating a systematic variance inflation.
We fit each data set to Eq.~\ref{eq: broadeningActuator} and list the resulting variance inflation factors, $\kappa$, in Table~\ref{tab: correlationSchemes}.
Because the finite-difference–induced correlations scale with the noise-driven actuator motion, their effect manifests as a multiplicative rescaling of the steady-state variance. 
As expected, $\kappa$ increases monotonically from Schemes~I to IV as the degree of temporal correlation is increased.
The resulting inflation factor $\kappa$ therefore depends on the specific modulation, demodulation, and update strategy, rather than being a universal property of the control loop.
In the present implementation, the sign-based control rule (Eq.~\ref{eq: controlLogic}) and single-difference demodulation (Eq.~\ref{eq: discErrorSignal}) impose a particular correlation structure. 
Modifying either, for example, by using proportional-step control or replacing differencing with a moving average, would alter the temporal correlations and hence the value of $\kappa$. 
Importantly, for white-noise inputs, $\kappa$ is independent of sampling frequency, since the correlations originate from the discrete-time error evaluation rather than the underlying noise spectrum.
\begin{table}[h!]
\centering
\caption{Representative modulation and update schemes illustrating the effect of correlated discriminator sampling on variance inflation.}
\label{tab: correlationSchemes}
\begin{tabular}{lllll}
\hline
\hline
\textbf{Sch.} & \textbf{Modulation} & \textbf{Update} & $\left<|\Delta \mu_{\rm a}|\right>/\nu_0$ (\%) & $\kappa$ \\
\hline
\rm{I} & Random RZ & Alternate Cycle & 0.06 &  1.000(1) \\
\rm{II} & Random NRZ & Every Cycle &  0.23 & 1.187(2) \\
\rm{III} & Random RZ & Every Cycle & 0.14 &  1.223(3) \\
\rm{IV} & Alternate NRZ & Every Cycle &  0.20 & 1.371(3) \\
\hline
\hline
\end{tabular}
\end{table}

\section{Colored Noise Regime: Breakdown of the Memoryless Approximation}\label{sec: coloredNoise}

\begin{figure}[t!]
\centering
\includegraphics[width=\linewidth]{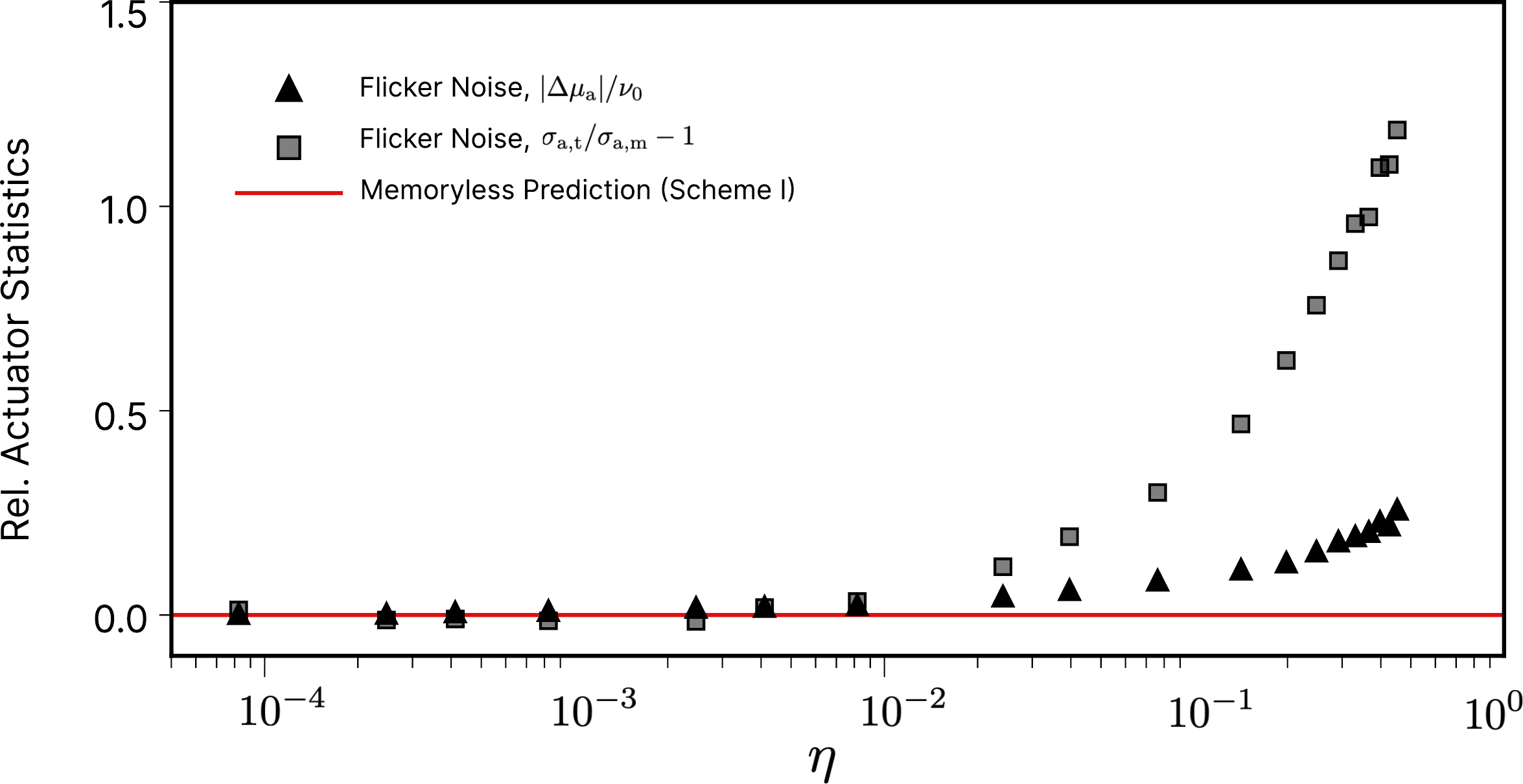}
\caption{Deviation between Markov-predicted and time-domain steady-state actuator statistics in the presence of colored frequency noise.
Normalized actuator mean shift, $|\Delta \mu_{\rm a}|/\nu_0$, and the relative variance mismatch, $\sigma_{\rm a,t}/\sigma_{\rm a,m}-1$, plotted as a function of the fractional flicker-noise contribution $\eta$.
The horizontal line indicates Markov prediction for the memoryless case.}
\label{fig: coloredNoise}
\end{figure}

Unlike white noise, which remains uncorrelated under discrete-time sampling, colored frequency noise introduces temporal correlations that span multiple control updates.
These correlations violate the strictly memoryless assumption underlying the Markov construction and render the actuator dynamics dependent on the sampling rate. 
As a result, the correspondence between Markov-predicted and the time-domain steady-state actuator distributions no longer hold.
In practical PIC lasers, where low-frequency flicker noise often dominates, such long-range correlations imprint additional structure on the discriminator signal and the resulting actuator dynamics.
To illustrate this, we perform a set of simulations where we progressively increase the flicker-noise strength, while fixing all other system parameters.
Specifically, we vary the flicker-noise strength parameter ${\rm S}_0$ in Eq.~\ref{eq: freqNoise} from $10^{5}$ to $10^{9}$ Hz$^3$/Hz, while fixing $\Delta \nu_{\rm lw} = 100$ kHz, $\sigma_{\rm d} = 10^{-5}~\textbf{D}_{\rm span}$, $\mathnormal{w} = 2.5$ MHz, and $\Delta \nu_{\rm a}= 5$ kHz. 
Increasing ${\rm S}_0$ simultaneously increases both the magnitude and the correlation time of the low-frequency noise.
For each ${\rm S}_0$, we compute both the normalized actuator mean shift $|\Delta \mu_{\rm a}|/\nu_0$ and the relative variance mismatch $\sigma_{\rm a,t}/\sigma_{\rm a,m}-1$, and plot them as a function of the fractional flicker-noise contribution
\begin{equation}
\eta = \frac{\int_a^{f_{\rm s}/2}~{\rm S}_0/f^\alpha~df}{\int_a^{f_{\rm s}/2}~{\rm S}_{\rm f}(f)~df},
\label{eq: eta}
\end{equation}
in Fig.~\ref{fig: coloredNoise}. 
Here $a = 1$ Hz.
As $\eta$ increases, both the actuator mean and variance depart systematically from the Markov prediction, with the mean assuming a non-zero bias and the variance exceeding the white-noise inflation factor $\kappa=1$.
These results demonstrate that colored frequency noise produces a correlation-driven distortion of the steady-state actuator distribution that cannot be captured by a memoryless Markov model or absorbed into a single variance correction factor.
The white-noise inflation factor therefore represents a lower bound on actuator variance, while low-frequency noise introduces additional, sampling-dependent bias and excess variance.
In the flicker-noise-dominated regime, this behavior reflects the growing influence of noise history on the discriminator response, motivating finite-memory extensions of the Markov description.

\section{Conclusion}\label{sec: conclusion}

We have presented a generalized, Markov-chain framework for analyzing digital laser frequency stabilization systems. 
By representing the quantized actuator as a discrete-state variable and deriving transition probabilities from discriminator errors evaluated independently at each control update, this approach provides a transparent and computationally efficient means of predicting key stability metrics, such as steady-state actuator distributions and locked-frequency statistics. 
The framework captures digital non-idealities, such as sensing and actuator quantization, measurement noise, and sampling delays, while also incorporating control loop design choices such as the dither scheme and the control logic.
In the white-noise regime, the augmented transition-probability formulation reproduces the actuator and locked-frequency variances obtained from full time-domain simulations, while avoiding the extensive time averaging and rare-event sampling required by time-domain analyses.
Because the steady state is obtained directly from the transition matrix, the computational cost scales primarily with the number of discrete actuator states rather than the simulated time horizon or noise correlation time. 
This structural advantage enables rapid and systematic exploration of control-loop designs, particularly in low-noise regimes where time-domain simulations converge slowly.

When low-frequency flicker noise dominates, the Markov framework reveals a systematic and predictable departure from the memoryless assumption. 
As the fractional contribution of flicker noise increases, long-range temporal correlations persist across control updates and progressively modify the effective transition statistics. 
This leads to a monotonic deviation of the actuator statistics beyond the white-noise baseline, as quantified by the increasing actuator mean shift, $\Delta \mu_{\rm a}$, and the variance ratio, $\sigma_{\rm a,t}/\sigma_{\rm a,m}$, with flicker fraction, $\eta$. 
Importantly, this behavior reflects the growing influence of noise history on the discriminator response when correlations extend across update cycles. 
In this regime, the white-noise variance inflation factor constitutes a lower bound, while colored noise introduces additional, correlation-driven variance that cannot be absorbed into a single sampling-independent correction factor. 
These results clarify the regime of validity of memoryless Markov descriptions and motivate finite-memory extensions that explicitly account for temporal correlations, while preserving the Markov framework as a useful and physically transparent baseline.

Capturing these correlations exactly would require augmenting the actuator state with a finite history of prior noise samples, resulting in a higher-order or finite-memory Markov model whose transition-matrix dimension grows from ${\rm N}$ to ${\rm N}^p$, where ${\rm N}$ is the number of discrete actuator states and $p$ is the memory length. 
Even modest memory lengths therefore lead to prohibitively large eigenproblems, making a full finite-order treatment impractical for routine design exploration. 
The observed breakdown thus motivates reduced-order extensions that capture the dominant effects of temporal correlations without sacrificing interpretability or computational efficiency. 
Approaches such as approximating the low-frequency noise component with a low-order autoregressive process and applying an associated variance-correction factor offer a path forward. 
Together, these results establish the augmented Markov-chain framework as a practical and unifying tool for digital laser frequency lock analysis, providing quantitative steady-state predictions, clear design guidance, and favorable computational scaling for PIC-based and multi-laser platforms.

{\appendices
\section{Numerical Realizations of Free-Running Laser Noise}
\label{apdx: laserNoise}

We generate time-domain realizations of the laser frequency noise, $\delta \nu$, such that its ensemble-averaged PSD follows Eq.~\ref{eq: freqNoise}. 
To generate these numerical realizations, we synthesize the composite noise spectrum in the frequency domain at a sampling frequency $f_{\rm s} = 1$ MHz, with ${\rm N}_{\rm f}$ points spanning the frequency range $[-f_{\rm s}/2,~+f_{\rm s}/2]$. 
To introduce stochasticity, we choose the white noise component such that its amplitude follows a Rayleigh density distribution with a scale parameter $\sigma_{\rm f}$ \cite{kuo2018, goodman2025}.
We determine $\sigma_{\rm f}$, by equating the Rayleigh distribution's second moment, $2 \sigma_{\rm f}^2$, to that of the desired white noise, according to
\begin{equation}
2 \sigma_{\rm f}^2 = \frac{{\rm N}_{\rm f} f_{\rm s} \Delta\nu_{\rm lw}}{2 \pi}.
\end{equation}
We implement the flicker component by scaling the spectral magnitude according to $\sqrt{{\rm S}_0/f^\alpha}$. 
Since the white- and flicker-noise components are statistically independent, we add their amplitudes in quadrature to obtain the total spectral magnitude for each frequency bin.
To avoid artificial correlations, we assign random phases uniformly distributed between 0 and 2$\pi$ to all non-negative frequency bins. 
After setting the magnitude and phase for the positive-frequency bins, we impose Hermitian symmetry by setting the negative-frequency bins to be complex conjugates of the corresponding positive-frequency bins, to guarantee that the inverse Fourier transform yields a real-valued time-domain signal.
The black line in Fig.~\ref{fig: laserNoise} shows the one-sided PSD, of a representative realization with $\Delta \nu_{\rm lw} = $ 100 kHz, ${\rm S}_0 = 10^{9}$ Hz$^3$/Hz, and ${\rm N}_{\rm f} \approx 10^{6}$.
This realization is consistent with the ensemble-averaged PSD given by Eq.~\ref{eq: freqNoise}. 
We denote the white noise level by drawing a horizontal dashed line corresponding to $\Delta\nu_{\rm lw}/\pi$. 
While the flicker component dominates the PSD for frequencies $f< 10$ kHz, the white noise dominates higher frequencies. 
The red points in Fig.~\ref{fig: laserNoise} correspond to a laser whose frequency is locked using a low-bandwidth digital lock, described later in Sec.\ref{sec: coloredNoise}.
We obtain the time-domain laser frequency noise, $\delta \nu$, by performing an inverse Fourier transform of the frequency-domain realization, which provides realistic inputs for examining how the laser’s stochastic fluctuations are processed by the digital discriminator and stabilized by the feedback loop.
\begin{figure}[t]
\centering
\includegraphics[width=\linewidth]{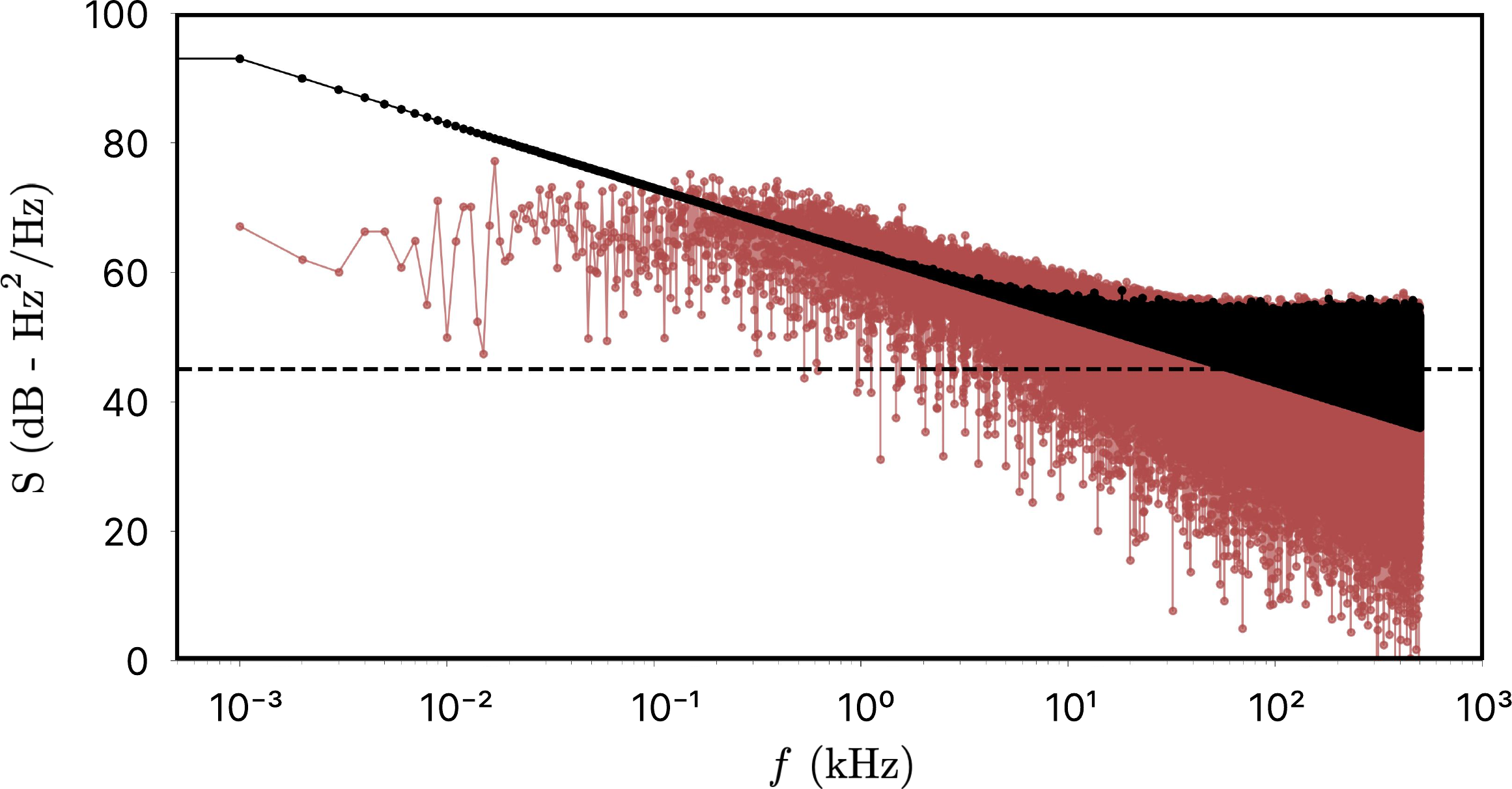}
\caption{Single instance of the free running ($S_{\rm f}$, black) and locked (red) laser's frequency noise spectra. The dashed horizontal line indicates the white noise level.}
\label{fig: laserNoise}
\end{figure}

\section{Representative Frequency Discriminator Responses}
\label{apdx: discriminator}

This appendix presents representative Voigt discriminator responses to illustrate how variations in sensor parameters affect the error signal slope and the resulting stochastic behavior of the digital feedback loop.
Fig.~\ref{fig: discriminator} shows two example discriminators with identical Lorentzian widths of $\gamma = 1$ MHz but different Gaussian widths, $\mathnormal{w} = 0.5$ MHz and 2.5 MHz, plotted as the purple and red curves, respectively.
Figures~\ref{fig: discriminator}(a) and (b) show the raw discriminator output $\mathbf{D}(\Delta\nu)$ and the corresponding mean demodulated error signal $\langle\widetilde{\mathbf{D}}\rangle(\Delta\nu)$, respectively.
For these examples, we set $\Delta \textbf{D} = \textbf{D}_{\rm span}/2^{12}$ and $\sigma_{\rm d} = 10^{-5}~\textbf{D}_{\rm span}$.
We estimate the mean demodulated error signal for an applied modulation $|\Delta {\rm M}| = 40$ kHz by averaging the positive and negative contributions according to $\langle\widetilde{\mathbf{D}}\rangle(\Delta \nu) = |\Delta {\rm M}|~\left(\mathbf{D}(\Delta \nu+|\Delta {\rm M}|) - \mathbf{D}(\Delta \nu-|\Delta {\rm M}|)\right)/2$.
Because the narrower, Lorentzian-dominated discriminator has a sharper effective linewidth, its error signal exhibits a correspondingly steeper slope than the Gaussian-broadened case. 
These slope differences directly influence both the sensitivity and the stochastic behavior of the digital feedback loop.
By capturing all discriminators with the parameters $\{ \gamma,~\mathnormal{w},~\Delta \textbf{D},~\sigma_{\rm d}\}$, the discretized Voigt model provides a unified framework for the Markov analysis, requiring only parameter substitution for different sensor types.
\begin{figure}[t]
\centering
\includegraphics[width=\linewidth]{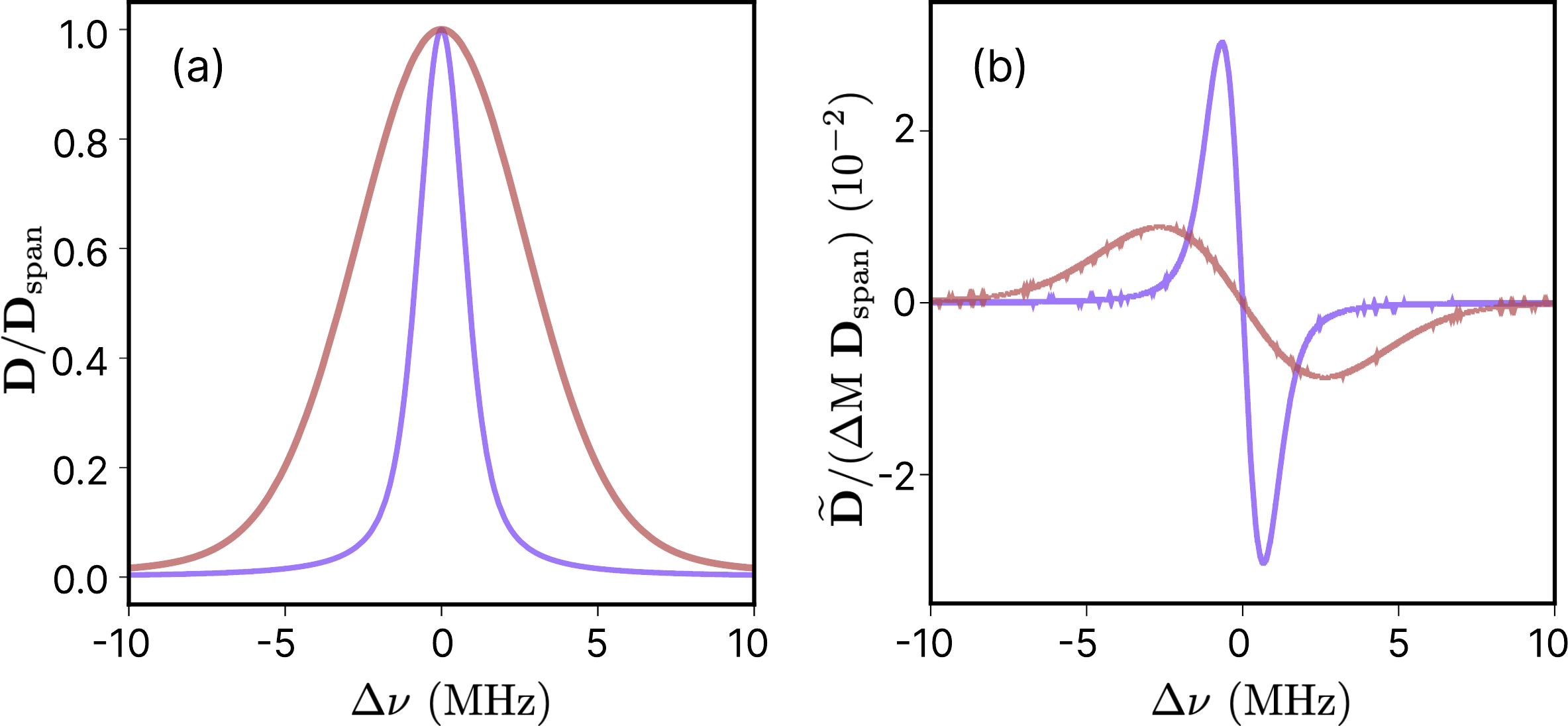}
\caption{Unified Voigt-based model for photonic frequency discriminators.
(a) Normalized discriminator response $\mathbf{D}(\Delta\nu)$ and (b) mean demodulated error signal $\langle\widetilde{\mathbf{D}}\rangle(\Delta\nu)$ plotted as a function of frequency detuning. 
The purple and red curves correspond to discriminators with the same Lorentzian width $\gamma = 1$ MHz but different Gaussian widths, $\mathnormal{w} = 0.5$ MHz and 2.5 MHz, respectively.}
\label{fig: discriminator}
\end{figure}
}

\bibliographystyle{IEEEtran}
\bibliography{sample}

@article{sun2020,
author = {Weiyang Sun and Zepeng Ning},
title = {Quantised output-feedback design for networked control systems using semi-Markov model approach},
journal = {International Journal of Systems Science},
volume = {51},
number = {9},
pages = {1637--1652},
year = {2020},
publisher = {Taylor \& Francis},
doi = {10.1080/00207721.2020.1772400},
URL = { https://doi.org/10.1080/00207721.2020.1772400},
eprint = { https://doi.org/10.1080/00207721.2020.1772400}
}

@article{zhang2023,
AUTHOR = {Zhang, Fan and Hua, Mingang and Gao, Mengyu},
TITLE = {Dynamic Output Feedback Quantization Control of a Networked Control System with Dual-Channel Data Packet Loss},
JOURNAL = {Mathematics},
VOLUME = {11},
YEAR = {2023},
NUMBER = {11},
ARTICLE-NUMBER = {2544},
URL = {https://www.mdpi.com/2227-7390/11/11/2544},
ISSN = {2227-7390},
DOI = {10.3390/math11112544}
}

@article{shen2018,
author={Shen, Mouquan
and Zhang, Hainan
and Park, Ju H.},
title={Observer-based quantized sliding mode ${H}_{\infty }$ control of Markov jump systems},
journal={Nonlinear Dynamics},
year={2018},
month={Apr},
day={01},
volume={92},
number={2},
pages={415-427},
issn={1573-269X},
doi={10.1007/s11071-018-4064-x},
url={https://doi.org/10.1007/s11071-018-4064-x}
}

@article{hsu2015,
author = {Hsu, Wen-Chiung and Lee, Lian-Wang and Tseng, Kuan-Hsuan and Lu, Chien-Yu and Liao, Chin-Wen and Shou, Ho-Nien},
title = {Design of Feedback Control for Networked Finite-Distributed Delays Systems with Quantization and Packet Dropout Compensation},
journal = {Discrete Dynamics in Nature and Society},
volume = {2015},
number = {1},
pages = {158972},
doi = {https://doi.org/10.1155/2015/158972},
url = {https://onlinelibrary.wiley.com/doi/abs/10.1155/2015/158972},
eprint = {https://onlinelibrary.wiley.com/doi/pdf/10.1155/2015/158972},
year = {2015}
}

@article{rønnekleiv2001,
title = {Frequency and Intensity Noise of Single Frequency Fiber Bragg Grating Lasers},
journal = {Optical Fiber Technology},
volume = {7},
number = {3},
pages = {206-235},
year = {2001},
issn = {1068-5200},
doi = {https://doi.org/10.1006/ofte.2001.0357},
url = {https://www.sciencedirect.com/science/article/pii/S1068520001903578},
author = {Erlend Rønnekleiv}
}

@article{martin-sanchez2024,
author = {David Martin-Sanchez and Edward Z. Zhang and Jake Paterson and James A. Guggenheim and Zhixin Liu and Paul C. Beard},
journal = {Opt. Lett.},
number = {3},
pages = {678--681},
publisher = {Optica Publishing Group},
title = {Laser frequency noise characterization using high-finesse plano\&\#x2013;concave optical microresonators},
volume = {49},
month = {Feb},
year = {2024},
url = {https://opg.optica.org/ol/abstract.cfm?URI=ol-49-3-678},
doi = {10.1364/OL.510516},
}

@Article{jin2021,
author={Jin, Warren
and Yang, Qi-Fan
and Chang, Lin
and Shen, Boqiang
and Wang, Heming
and Leal, Mark A.
and Wu, Lue
and Gao, Maodong
and Feshali, Avi
and Paniccia, Mario
and Vahala, Kerry J.
and Bowers, John E.},
title={Hertz-linewidth semiconductor lasers using CMOS-ready ultra-high-Q microresonators},
journal={Nature Photonics},
year={2021},
month={May},
day={01},
volume={15},
number={5},
pages={346-353},
issn={1749-4893},
doi={10.1038/s41566-021-00761-7},
url={https://doi.org/10.1038/s41566-021-00761-7}
}

@ARTICLE{antona2022,
  author={Antona, Gaspare and Gozzelino, Michele and Micalizio, Salvatore and Calosso, Claudio E. and Costanzo, Giovanni A. and Levi, Filippo},
  journal={IEEE Transactions on Instrumentation and Measurement}, 
  title={Frequency Noise Characterization of Diode Lasers for Vapor-Cell Clock Applications}, 
  year={2022},
  volume={71},
  number={},
  pages={1-8},
  doi={10.1109/TIM.2022.3216836}}

@article{
Guo2022,
author = {Joel Guo  and Charles A. McLemore  and Chao Xiang  and Dahyeon Lee  and Lue Wu  and Warren Jin  and Megan Kelleher  and Naijun Jin  and David Mason  and Lin Chang  and Avi Feshali  and Mario Paniccia  and Peter T. Rakich  and Kerry J. Vahala  and Scott A. Diddams  and Franklyn Quinlan  and John E. Bowers },
title = {Chip-based laser with 1-hertz integrated linewidth},
journal = {Science Advances},
volume = {8},
number = {43},
pages = {eabp9006},
year = {2022},
doi = {10.1126/sciadv.abp9006},
URL = {https://www.science.org/doi/abs/10.1126/sciadv.abp9006},
eprint = {https://www.science.org/doi/pdf/10.1126/sciadv.abp9006}}

@article{stern2020,
author = {Liron Stern and Wei Zhang and Lin Chang and Joel Guo and Chao Xiang and Minh A. Tran and Duanni Huang and Jonathan D. Peters and David Kinghorn and John E. Bowers and Scott B. Papp},
journal = {Opt. Lett.},
number = {18},
pages = {5275--5278},
publisher = {Optica Publishing Group},
title = {Ultra-precise optical-frequency stabilization with heterogeneous III--V/Si lasers},
volume = {45},
month = {Sep},
year = {2020},
url = {https://opg.optica.org/ol/abstract.cfm?URI=ol-45-18-5275},
doi = {10.1364/OL.398845},
}

@book{yariv2007, 
place={Oxford, UK}, edition={6th}, 
title={Photonics: Optical Electronics in Modern Communications}, 
ISBN={9780195179460}, 
abstractNote={The field of photonics, sometimes referred to as optical electronics, has continued to evolve vigorously during the last decade, thus justifying a major updating of the last (fifth) edition. The present edition has a broader theoretical underpinning and includes new and important subjects. The book continues the tradition of introducing basic principles in a systematic self- contained treatment with minimal reliance on outside sources. It describes the physics and methodology of operation of the basic optoelectronic components of importance to optical communications and optical electronics. The book is intended to serve both as a text for students in electrical engineering and applied physics as well as a reference book for engineers and scientists working in those fields. The present edition reflects two major efforts on our part: (1) the addition of new topics related to technology development in optical electronics and communications (and the omission of some less important topics) and (2) the refinement and improvement of materials already in the fifth edition. In the revision process, we decided to tailor the new edition for students, researchers, and engineers in the area of optical communications who are interested in learning how to generate and manipulate optical radiation and how to put this knowledge to work in analyzing and designing photonic components for the transmission of information. The presentation and inclusion of topics also reflect comments and suggestions from many anonymous reviewers and instructors. Specifically, the main new features of this edition are: 1. The introduction of Stokes parameters and the Poincaré sphere for the representation of polarization states in birefringent optical networks. 2. The use of Fermat's principle for the derivation of rays, beam propagation, and the Fresnel diffraction integral. 3. The use of matrix methods for treating wave propagation in coupled resonator optical waveguides (CROWs). 4. Matrix treatment of multicavity etalons and multilayer structures. 5. Matrix treatment of mode coupling and supermodes in mode-locked lasers. 6. Chromatic dispersion, polarization mode dispersion (PMD) in fibers, and their compensation. 7. Nonlinear optical effects in fibers: self-phase modulation, cross-phase modulation, stimulated Brillouin scattering (SBS) and stimulated Raman scattering (SRS) in fibers, optical four-wave mixing, and spectral reversal (phase conjugation) in fibers. 8. Electroabsorption and waveguide electro-optic Mach-Zehnder modulators. 9. Periodic layered media, fiber Bragg gratings and photonic crystals, and Bragg reflection waveguides. 10. Optical amplifiers: semiconductor optical amplifiers, erbium-doped fiber amplifiers, and Raman amplifiers. As in the earlier editions, we assume a basic background in electromagnetic theory and familiarity with Maxwell's equations and electromagnetic wave propagation in the bulk and in waveguides. An elementary acquaintance with quantum mechanics is recommended but may be acquired en route. A generous use of numerical examples is intended to help bridge the gap between theory and applications.}, publisher={Oxford University Press}, author={Yariv, Amnon and Yeh, Pochi}, year={2007}, pages={850} }

@ARTICLE{little1997,
  author={Little, B.E. and Chu, S.T. and Haus, H.A. and Foresi, J. and Laine, J.-P.},
  journal={Journal of Lightwave Technology}, 
  title={Microring resonator channel dropping filters}, 
  year={1997},
  volume={15},
  number={6},
  pages={998-1005},
  doi={10.1109/50.588673}}

@BOOK{okamoto2014,
  title     = "Fundamentals of optical waveguides",
  author    = "Okamoto, Katsunari",
  publisher = "Academic Press",
  edition   =  2,
  month     =  may,
  year      =  2014
}

@article{black2001,
    author = {Black, Eric D.},
    title = {An introduction to Pound–Drever–Hall laser frequency stabilization},
    journal = {American Journal of Physics},
    volume = {69},
    number = {1},
    pages = {79-87},
    year = {2001},
    month = {01},
    issn = {0002-9505},
    doi = {10.1119/1.1286663},
    url = {https://doi.org/10.1119/1.1286663},
    eprint = {https://pubs.aip.org/aapt/ajp/article-pdf/69/1/79/10115998/79_1_online.pdf},
}

@Article{drever1983,
author={Drever, R. W. P.
and Hall, J. L.
and Kowalski, F. V.
and Hough, J.
and Ford, G. M.
and Munley, A. J.
and Ward, H.},
title={Laser phase and frequency stabilization using an optical resonator},
journal={Applied Physics B},
year={1983},
month={Jun},
day={01},
volume={31},
number={2},
pages={97-105},
issn={1432-0649},
doi={10.1007/BF00702605},
url={https://doi.org/10.1007/BF00702605}
}

@book{rubiola2008, 
place={Cambridge}, series={The Cambridge RF and Microwave Engineering Series}, title={Phase Noise and Frequency Stability in Oscillators}, publisher={Cambridge University Press}, author={Rubiola, Enrico}, year={2008}, collection={The Cambridge RF and Microwave Engineering Series}}

@book{feller1968,
  title={An Introduction to Probability Theory and Its Applications, Volume 1},
  author={Feller, William},
  year={1968},
  publisher={John Wiley \& Sons},
  address={New York},
  edition={3rd},
  series={Wiley Series in Probability and Statistics},
  pages={xviii+509},
  isbn={9780471257080}
}

@book{ross2020,
  author    = {Ross, Sheldon M.},
  title     = {Introduction to Probability Models},
  edition   = {12th},
  publisher = {Academic Press (Elsevier)},
  year      = {2020},
  isbn      = {978-0128143469},
  address   = {London, UK},
  note      = {Includes new coverage on coupling, renewal theory, queueing theory, and Poisson processes; winner of a 2020 Textbook Excellence Award (Texty)}
}

@book{norris97,
  address = {Cambridge, UK},
  author = {Norris, J. R.},
  publisher = {Cambridge University Press},
  title = {Markov Chains},
  year = 1997
}

@Article{idjadi2024,
author={Idjadi, Mohamad Hossein
and Kim, Kwangwoong
and Fontaine, Nicolas K.},
title={Modulation-free laser stabilization technique using integrated cavity-coupled Mach-Zehnder interferometer},
journal={Nature Communications},
year={2024},
month={Mar},
day={01},
volume={15},
number={1},
pages={1922},
issn={2041-1723},
doi={10.1038/s41467-024-46319-3},
url={https://doi.org/10.1038/s41467-024-46319-3}
}

@Article{zhang2024,
author={Zhang, Xuguang
and Zhou, Zixuan
and Guo, Yijun
and Zhuang, Minxue
and Jin, Warren
and Shen, Bitao
and Chen, Yujun
and Huang, Jiahui
and Tao, Zihan
and Jin, Ming
and Chen, Ruixuan
and Ge, Zhangfeng
and Fang, Zhou
and Zhang, Ning
and Liu, Yadong
and Cai, Pengfei
and Hu, Weiwei
and Shu, Haowen
and Pan, Dong
and Bowers, John E.
and Wang, Xingjun
and Chang, Lin},
title={High-coherence parallelization in integrated photonics},
journal={Nature Communications},
year={2024},
month={Sep},
day={10},
volume={15},
number={1},
pages={7892},
issn={2041-1723},
doi={10.1038/s41467-024-52269-7},
url={https://doi.org/10.1038/s41467-024-52269-7}
}

@BOOK{goodman2025,
  title     = "Speckle phenomena in optics",
  author    = "Goodman, Joseph W",
  publisher = "SPIE Press",
  series    = "Press Monographs",
  edition   =  2,
  month     =  may,
  year      =  2025,
  address   = "Bellingham, WA",
  language  = "en"
}

@INCOLLECTION{kuo2018,
  title     = "Delta Functions",
  booktitle = "White Noise Distribution Theory",
  author    = "Kuo, Hui-Hsiung",
  publisher = "CRC Press",
  pages     = "61--76",
  month     =  may,
  year      =  2018
}

@book{levin2006,
  author = {Levin, David A. and Peres, Yuval and Wilmer, Elizabeth L.},
  publisher = {American Mathematical Society},
  title = {{Markov chains and mixing times}},
  url = {http://scholar.google.com/scholar.bib?q=info:3wf9IU94tyMJ:scholar.google.com/&output=citation&hl=en&as_sdt=2000&ct=citation&cd=0},
  year = 2006
}

\end{document}